\documentclass[11pt, a4paper]{article}
\usepackage{jheppub}
\usepackage{graphicx}
\usepackage{fancyvrb}
\usepackage{makeidx}
\usepackage{latexsym,amssymb,amsmath,graphicx, makeidx,slashed}
\usepackage{epstopdf}

\newcommand{\beq}{\begin{equation}}
\newcommand{\beqa}{\begin{eqnarray}}
\DefineVerbatimEnvironment{code}{Verbatim}{fontsize=\small}
\newcommand{\eeq}{\end{equation}}
\newcommand{\eeqa}{\end{eqnarray}}
\newcommand{\bit}{\begin{itemize}}
\newcommand{\eit}{\end{itemize}}
\newcommand{\cL}{{\cal L}}
\newcommand{\cO}{{\cal O}}
\newcommand{\cA}{{\cal A}}
\newcommand{\cc}{{\rm c.c}}

\makeindex

\title{
A Supersymmetric Higgs Sector with Chiral $D$-terms}

\author[a,b]{Nathaniel Craig,}
\author[c]{and Andrey Katz}
\affiliation[a]{Department of Physics, Rutgers University, Piscataway, NJ 08854}
\affiliation[b]{School of Natural Sciences, Institute for Advanced Study, Princeton, NJ 08540}
\affiliation[c]{Center for the Fundamental Laws of Nature, Jefferson Physical Laboratory, Harvard University, Cambridge, MA 02138}

\abstract{Although supersymmetry remains the best candidate for solving the electroweak hierarchy problem,
a supersymmetric Higgs boson near 125 GeV requires  heavy scalars, highly-mixed stops, or non-minimal contributions
to the Higgs potential. Extensions of the Standard Model (SM) gauge group provide an attractive means of raising the
Higgs mass through non-decoupling $D$-term contributions to the Higgs quartic, but in most cases this correction is
correlated with an enhanced coupling to bottom quarks and tau leptons that is disfavored by current fits to 
LHC Higgs data. In this
work we demonstrate that the Higgs mass may be raised by non-decoupling $D$-terms without such 
enhanced couplings if the two supersymmetric Higgs doublets are ``chiral'', i.e., charged under different 
gauge groups at high
energies. In this case there is no direct correlation between the correction to the Higgs mass and its 
couplings to SM states, and in
general the chiral correction to the Higgs potential undoes the MSSM preference for enhanced bottom couplings. This raises the prospects 
for discovering additional supersymmetric Higgs bosons consistent with the measured mass and couplings of the observed Higgs. 
}

\preprint{RUNHETC-2012-23}
\begin{document}
\maketitle

\section{Introduction}
The recent discovery of a Higgs-like resonance~\cite{Atlas:higgsdisc:2012gk,CMS:higgsdisc:2012gu}
and the first measurements of its branching ratios (BRs) provide 
a litmus test for models for physics beyond the Standard Model (SM). The massive vector decay modes 
$h \to WW^*$ and $h \to ZZ^*$ have been measured and agree well with the predictions for a SM-like Higgs. 
At the same time, hints of possible deviations from the SM predictions in the $h \to \gamma \gamma$ channel 
are fairly inconclusive and may well prove to be statistical fluctuations. 
The current fit of the Higgs BRs to the SM looks reasonable for the time being, strongly indicating that the 
Higgs is essentially SM-like and the deviations from the SM predictions
are likely small.

The mass of the Higgs is somewhat perplexing from the point of view of new physics. 
While it is probably  too light to be a  ``partially composite'' Higgs, or to fit nicely within various generic 
technicolor-inspired models,  it is also slightly too heavy to naturally fit the paradigm of minimal 
supersymmetry (SUSY). The minimal supersymmetric standard model (MSSM) at tree level predicts
\beq
m_h \leq m_Z |\cos (2\beta)|
\eeq
due to the supersymmetric relation between the Higgs quartic and the electroweak gauge couplings.
Radiative corrections dominated by stop-top loops can push the Higgs mass somewhat beyond this bound, but 
the corrections grow only logarithmically with the stop mass. A Higgs mass $\sim 125$ GeV requires stop masses 
of order 10 TeV, largely undermining the very motivation of SUSY as a solution to the hierarchy problem of the SM.
Alternately, the Higgs mass may be raised by finite threshold corrections due to large stop mixing,
but this requires a considerable numerical conspiracy between soft masses  that is hard to achieve in many models of SUSY breaking. For a summary of the implications for the MSSM, see e.g.~\cite{Hall:2011aa, Heinemeyer:2011aa, Arbey:2011ab, Arbey:2011aa, Draper:2011aa, Carena:2011aa, Cao:2012fz,  Christensen:2012ei, Brummer:2012ns}.

Supersymmetric naturalness -- and the prospects for discovering additional states at the LHC -- may be 
preserved in light of the observed Higgs mass if there are extra contributions to the Higgs quartic coupling 
above and beyond those of the MSSM. An effective field theory
approach to such models was developed in~\cite{Dine:2007xi}, and concrete examples for these
models include, among other possibilities, additional $SU(2)$  triplets~\cite{Espinosa:1991gr,Espinosa:1992hp}; $SU(2)$ 
doublets~\cite{Azatov:2011ht, Azatov:2012wq}; or gauge singlets (see~\cite{Ellwanger:2009dp}
for review) that couple to the MSSM Higgses;  or some combination thereof~\cite{Agashe:2011ia}. 
While none of these explanations is firmly excluded, in general 
they modify the Higgs branching ratios compared to the SM. It is challenging to find a parameter space for 
these models in which the lightest CP-even scalar masquerades effectively as the SM Higgs without any 
noticeable deviation from the measured
branching ratios.

An alternate possibility is to enhance the SUSY Higgs quartic through additional non-decoupling $D$-terms, 
as first introduced in~\cite{Batra:2003nj}. Such $D$-terms require a gauge extension of the MSSM group slightly 
above the TeV scale. For example, consider an $SU(2)\times SU(2)$ gauge group broken to a diagonal
subgroup that we identify as the $SU(2)_L$ weak gauge group of the SM. If the MSSM
Higgs doublets are originally charged under one of the $SU(2)$'s at high energies,
they will have a $D$-term potential coming from both the unbroken SM gauge group and from the heavy $W', Z'$ gauge fields of the broken 
group. Such models were studied in detail in~\cite{Batra:2003nj,Maloney:2004rc, Medina:2009ey} and found to possess various attractive properties, 
including almost-vanishing contributions to electroweak precision measurements and preservation of gauge coupling unification. Unfortunately these
models also change the Higgs branching ratios, particularly the partial width $\Gamma(h \to b \bar b)$. As shown in~\cite{Blum:2012ii}, these models are already in some tension with experimental data, which has only grown stronger as Higgs couplings are measured more accurately. While the enhancement of $\Gamma(h \to b \bar b)$ may be decoupled by raising the masses of other scalars in the Higgs sector, this erases the prospects for 
discovering additional Higgs scalars at the LHC.  This tension suggests $D$-term contributions to the potential, if responsible for raising the Higgs mass, should take an unconventional form.

In this work we will focus on non-decoupling $D$-terms that come from a conceptually
different ultraviolet source, in which the MSSM Higgs doublets are chiral at high energies. A model along these lines was proposed
in~\cite{Craig:2011yk}, but the full implications of this scenario for the Higgs sector
remain unexplored. In this work we examine the effects of the non-decoupling chiral $D$-terms on the Higgs 
sector. We show that, unlike in the ``standard'' scenario of non-decoupling $D$-terms,
corrections from the extended gauge sector tend to \emph{suppress} the $h \to b \bar b$ and
$h \to \tau^+ \tau^-$ rates relative to the MSSM expectation.
Since these rates are enhanced in the MSSM relative to the SM in the regime of large 
$\tan \beta $~\cite{Azatov:2012wq},
the chiral non-decoupling terms partially undo the ``damage'' to the Higgs potential caused by the MSSM itself.
We also emphasize that in spite of the naive  expectation, these chiral Higgs $D$-term can provide a good fit to precision electroweak measurements. In fact, the situation in the low-energy effective theory is improved compared to the MSSM.

The chiral Higgs scenario possesses a handful of non-trivial features. First, this scenario
provides a natural resolution to the $\mu/B\mu$ problem. The conventional $\mu$ term is forbidden by gauge invariance in the full
model, and only emerges beneath the scale of symmetry breaking to the SM gauge group. Second, the $B\mu$
term can typically be very small, favoring very large values for $\tan \beta$. In general, $\tan \beta $ can vary from $\sim 50 $
all the way to $\cO(10^4)$, where one-loop non-holomorphic couplings dominate down-type quark and lepton 
fermion masses~\cite{Borzumati:1999sp,Dobrescu:2010mk}. When $\tan \beta \sim 50$ and tree-level couplings dominate, the corrections 
to Higgs couplings are unambiguous. At larger values of $\tan \beta$ where loop-suppressed non-holomorphic couplings dominate, 
there may be additional corrections to Higgs couplings with ambiguous signs.

Our paper is organized as follows: In Sec.~\ref{sec:vectorlike} we review the standard non-decoupling
$D$-term scenario and discuss its problem with $b \bar b$  and $\tau^+ \tau^-$ Higgs decay modes. In Sec.~\ref{sec:chiral} we introduce chiral
Higgs $D$-terms and show that in the regime of moderately large $\tan \beta$ the corrections to the Higgs couplings fit SM
expectations \emph{better than the MSSM}.
We also illustrate that the regime of very large $\tan \beta $ regime, namely $\tan \beta \sim \cO(10^3 \ldots 10^4)$, is well motivated
in this scenario. We show that the Higgs branching ratios in this case do not significantly deviate from
the SM predictions  and, unlike in the moderate $\tan \beta $ regime, the corrections to the SM rate do not have a well-defined sign, but rather depend on various phases between soft terms.  Finally,
in Sec.~\ref{sec:conclusions} we conclude and suggest several other interesting questions related to this unusual ``chiral'' extension of the MSSM.

\section{Non-decoupling $D$-terms with vector Higgses}
\label{sec:vectorlike}
We begin by reviewing the implications of conventional non-decoupling $D$-terms wherein both MSSM Higgs doublets are charged under the same gauge group at high energies. Consider a gauge extension of the SM, such that the full gauge group above the scale $\sim 10$~TeV is
$SU(3)_c\times SU(2)_1\times SU(2)_2 \times U(1)_1\times U(1)_2$. It is also possible to double the $SU(3)$ group, in which case there are implications for mass mixing among colored sfermions. In this work we will not consider the $SU(3)\times SU(3)$ moose in detail since it is irrelevant for the Higgs discussion, but it is worth keeping this possibility in mind.  

In addition to MSSM fields, which may be charged under various choices of these groups, there are bi-fundamental link fields $\chi, \tilde \chi$. The pairs of $SU(2)$ and $U(1)$ gauge groups are
broken to their diagonal subgroups by VEVs of the link fields, which are in vector-like representations of
 $SU(2)_1 \times U(1)_1 \times SU(2)_2 \times U(1)_2$ and singlets of the color group, namely
\beq\label{eq:links}
\chi \sim (2_{1/2}, 2_{-1/2}), \ \  \ \ \tilde \chi \sim (2_{-1/2}, 2_{1/2}).
\eeq
The low-energy theory consists of the massless gauge fields of $SU(3)_c \times SU(2)_L \times U(1)_Y$, along with various massive states at the scale of the link field vevs. 

The breaking of two replicas of the electroweak gauge group to the diagonal subgroup can be easily achieved, and may be largely supersymmetric. For example, consider the superpotential
\beq \label{superpot}
W = \kappa \cA (\chi \tilde \chi - f^2)
\eeq
where $\cA$ is a gauge singlet dynamical field and $\kappa$ is a dimensionless Yukawa coupling . 
This structure is particularly motivated by strong SUSY
dynamics~\cite{Green:2010ww}. The fields $\chi,\ \tilde \chi$ and $\cA$ also obtain soft masses from SUSY breaking. Provided these soft masses are $\lesssim f$, the superpotential~\eqref{superpot} triggers VEVs for $\chi,\ \tilde \chi$ and breaks the extended gauge
group down to $SU(2)\times U(1)$, which we identify with the electroweak group at low energies.\footnote{Note that when SUSY is broken, in addition to quadratic soft terms we can also imagine a trilinear soft term
$\cL \supset a \cA \chi \tilde \chi + \cc$. This term would produce a tadpole for $\cA$, shifting it from the origin
and inducing non-vanishing $F$-terms for $\chi$ and $\tilde \chi$. In this particular analysis we have neglected this
term because we are motivated by gauge-mediation like scenarios where the $D$-terms are usually small and
therefore $a \ll \tilde m_{\chi}$. While there is nothing wrong with large $F_\chi$ in the Higgs analysis presented here, it can be more problematic when we try to embed it into a full theory that addresses the flavor
puzzle. The situation can become even worse in Sec.~\ref{sec:chiral}, where a large
$F$-term can induce an enormous $B\mu$ term and destabilize electroweak symmetry breaking.
Therefore it is important to focus on UV completions in which this $a$-term is not too large. }

Although the low-energy gauge theory is that of the MSSM,  novel phenomenology arises when MSSM fields are divided non-trivially among the gauge groups in the ultraviolet. 
Models along these lines~\cite{Craig:2011yk, Auzzi:2011eu, Craig:2012hc} can at least partially explain the flavor
puzzle of the SM, are attractive from the perspective of FCNCs~\cite{Barbieri:2011ci}, and lead to a ``natural'' SUSY spectrum~\cite{Dimopoulos:1995mi,Cohen:1996vb},
making it non-trivial for LHC discovery and opening a window for SUSY particles below
the TeV scale~\cite{Kats:2011qh,Brust:2011tb,Papucci:2011wy}.

Whatever the configuration of matter fields, there is a potential correction to $D$-terms of states charged under the electroweak gauge group at low energies due to the massive gauge bosons. Unsurprisingly, this correction is proportional to supersymmetry breaking; if the higgsing of gauge groups is perfectly supersymmetric, the low-energy $D$-terms are those of the MSSM in accordance with supersymmetric decoupling theorems. However, in the presence of SUSY-breaking soft terms for the link fields, the integrating-out is no longer supersymmetric and the following hard SUSY breaking terms are induced in the effective potential:
\beq
V \supset \frac{g^2 \Delta}{8}\left( H_u^\dagger \sigma^a H_u + H_d^\dagger \sigma^a H_d \right)^2 +
\frac{{g'}^2 \Delta'}{8} \left( H_u^\dagger H_u - H_d^\dagger H_d \right)^2~,
\eeq
where we have defined
\beq
\Delta \equiv \left( \frac{g_1}{g_2} \right)^2 \frac{M_s^2}{ M_V^2 + M_s^2}
\eeq
and a similar definition for $\Delta'$, with substitutions $g \to g'$ and $M_V \to M_V'$.  Here $M_s$
stands for the soft mass of the sgoldstone mode of the link fields and $M_V,\ M_V'$ are the masses of  heavy
$W'$ and $B'$ gauge bosons. As expected, these corrections are proportional to SUSY breaking and vanish in the SUSY limit.

This is a relatively clean and safe way to enhance the quartic coupling of the Higgs fields, which in turn raises the Higgs mass above
the MSSM bound. In this case we obtain a new tree-level bound on the Higgs mass,
\beq\label{eq:symmetricbound}
{m_h}^2_{tree} \leq \left( m_Z^2 + \frac{g^2 \Delta + {g'}^2 \Delta' }{2} v^2 \right) \cos^2 (2\beta)
\eeq
where $v \approx 174$ GeV. In order to match the observed Higgs mass without requiring large radiative corrections, we require (in the large $\tan \beta$
limit)
 \beq
\frac{g^2\Delta +{g'}^2 \Delta'}{2} v^2 \sim \cO(m_W^2)~.
\eeq

Unfortunately, this beautiful mechanism of addressing the Higgs mass problem in supersymmetry has unintended consequences for the Higgs branching ratios. Let us focus on the large $\tan \beta $ limit, which is necessary  to saturate the tree level bound on
the Higgs mass from~\eqref{eq:symmetricbound}. A generic prediction of these non-decoupling $D$-terms at large
$\tan \beta$ is an enhanced coupling of the Higgs to $b \bar b$ and $\tau^+ \tau^-$
without significant modification of other important Higgs couplings such as  $VV$ and $t \bar t$. To show this, we very
briefly review here relevant parts of a detailed analysis performed in~\cite{Blum:2012kn,Blum:2012ii}. Similar conclusions were 
obtained in~\cite{Azatov:2012wq}.

The MSSM Higgs sector is nothing but a two-Higgs-doublet model (2HDM) with a restricted set of parameters. The most generic renormalizable couplings of a 2HDM may be parameterized by the potential ~\cite{Gunion:2002zf}:\footnote{Note that this
reference uses 2HDM conventions with equal hypercharge for the two doublets; we prefer opposite-hypercharge conventions for a SUSY analysis. One can simply switch from our conventions to conventions of~\cite{Gunion:2002zf} by identifying
$\Phi_2 = H_u, \ \Phi_1 = i\sigma^2 H_d^*$.}
\beqa\label{eq:2hdm}
V & = & m_1^2 |H_d|^2 + m_2^2 |H_u|^2 +
\frac{\lambda_1}{2} |H_d|^4 + \frac{\lambda_2}{2} |H_u|^4 +
\lambda_3 |H_u|^2 |H_d|^2 +\lambda_4 |H_d^\dagger H_u|^2 + \\ \nonumber
&& \left( \frac{\lambda_5}{2}(H_u H_d)^2 + m_{12}^2 H_u H_d  + \lambda_6 |H_d|^2(H_u H_d) +
\lambda_7 |H_u|^2 (H_u H_d) + \cc \right).
\eeqa
 In the decoupling limit with large $\tan \beta$, the heavy states are identified with $H_d$; we can simply integrate this field out to obtain the couplings of fermions to the SM-like Higgs. If we neglect potential non-holomorphic couplings of
the Higgs to $b \bar b$ and $\tau^+ \tau^-$, we find the following expressions for the Higgs couplings to the
down-type fermions:
\beq
r_{b/\tau} \equiv \frac{g_{hb \bar b/h \tau^+ \tau^-}}{g_{h b \bar b/h \tau^+ \tau^-}^{SM}}
\approx \left( 1 + \frac{m_h^2 }{m_H^2} \right)
\left( 1 - \frac{2 (\lambda_3 + \lambda_5 )v^2 }{m_H^2 -m_h^2}\right) + \ldots
\eeq
where the ellipses stand  for terms that are usually negligible in SUSY models by virtue of being proportional
to $\lambda_7$ or non-holomorphic couplings.  The couplings to $b\bar b$ and $\tau^+ \tau^-$ are the only
Higgs couplings  that are corrected at the leading order at tree level in the limit of large $\tan \beta$. Corrections to $VV$ and $t \bar t$
couplings are
suppressed by $1/\tan^2 \beta$ and may be neglected. 

In the MSSM with non-decoupling $D$-terms, the correction to the $b \bar b$ and $\tau^+ \tau^-$ couplings takes the form
\beq
r_{b/\tau} \approx \left( 1+ \frac{m_h^2}{m_H^2} \right)\left(1+
\frac{(g^2 (1+ \Delta) + {g'}^2 (1+\Delta'))v^2}{2 (m_H^2 - m_h^2)} \right)~.
\eeq
Note that MSSM {\it without} any $D$-term corrections  (i.e., $\Delta = \Delta' = 0$) already enhances the rates for $h \to b \bar b$ and $h \to \tau^+ \tau^-$.
The corrections to the Higgs  quartic from non-decoupling $D$-terms further enhance
these rates by an amount proportional to the tree-level contribution to the Higgs mass. This result is fairly generic, and although it is not yet excluded, it is somewhat in odds with the data. In particular, such enhancement is constrained by direct measurements of  $h \to \tau^+ \tau^-$ and $h \to b \bar b$, as well as indirect limits on $h \to b \bar b$ coming from the measurement of Higgs decays to $WW^*$, $ZZ^*$ and $\gamma \gamma$. This is illustrated in Figure 1, which shows the tree-level Higgs mass prediction and the enhancement of $r_b$ relative to the Standard Model value.

\begin{figure}[htbp] 
   \centering
    \includegraphics[width=3in]{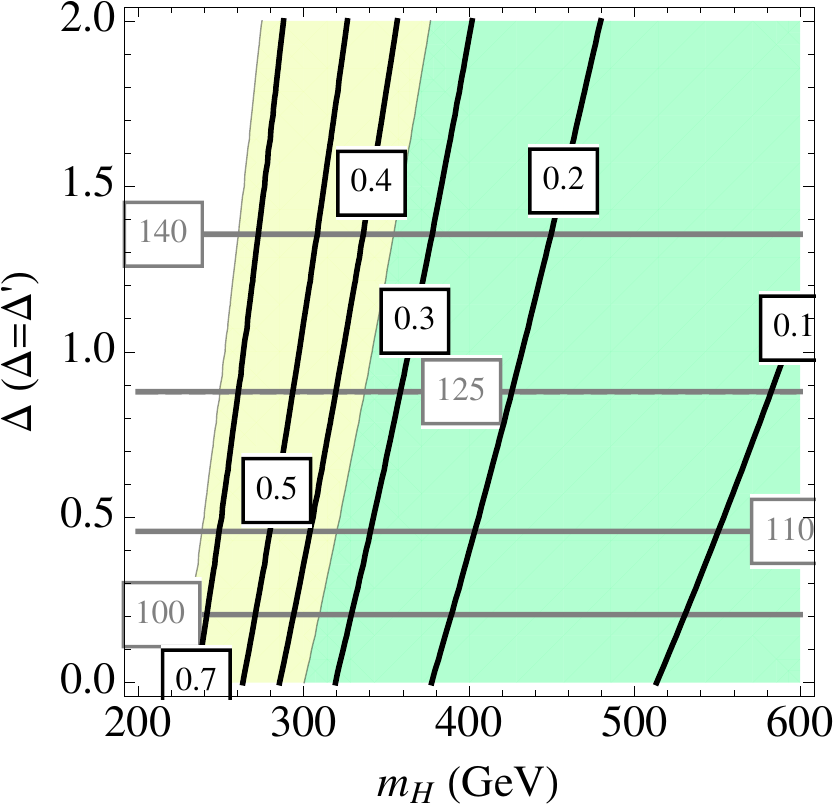} 
   \caption{The tree-level prediction for the Higgs mass in GeV (gray lines) and the enhancement of the bottom coupling $r_b - 1$ above the Standard Model value (black lines) in the vector case as a function of the parameter $\Delta$ (taking $\Delta = \Delta'$) and the heavy CP-even Higgs mass $m_H$. Additional radiative contributions from stop loops further raise the Higgs mass, but are not required to accommodate a Higgs mass around 125 GeV. The green region contains the $+ 1 \sigma$ fit to $r_b$ using the most recent CMS results~\cite{CMS-PAS-HIG-12-045}; yellow contains the $+2 \sigma$ fit. Points in the white region lie more than $2 \sigma$ outside the CMS fit to $r_b$. }
   \label{fig:vec}
\end{figure}

We can make this somewhat more quantitative by studying the current best fits to Higgs signals. Recently this 
has become a fairly precise procedure conducted directly by the collaborations, with full benefit of information 
about correlations between measurements in various channels. For simplicity we will focus our discussion on 
results from CMS~\cite{CMS-PAS-HIG-12-045}, which to date present the most comprehensive information on 
Higgs coupling fits using $5 \oplus 12$~fb$^{-1}$ of $7 \oplus 8$~TeV data.\footnote{For a theory-level fit 
focusing on coupling to down-type quarks, see~\cite{Azatov:2012wq}.} In a generic search for deviations in the 
six measured couplings (tree-level couplings to top quarks, bottom quarks, tau leptons, and vectors, plus 
effective couplings to gluons and photons), the 1D fit to the bottom quark coupling obtained by profiling the other 
five couplings yields $r_b \simeq 1.1^{+ 0.6}_{-0.5}$, while a similar fit to the tau coupling yields 
$r_\tau \simeq 0.95\pm 0.4$.  Although these coupling fits do not assume the particular relations among parameters 
present in the MSSM, they imply that current data is inconsistent with significant enhancement of couplings to 
down-type quarks. Similarly, in a search for deviations in the equivalence of couplings to down-type and 
up-type quarks (varying the coupling to vectors, the coupling to up-type quarks, and the coupling modifier 
ratio $\lambda_{du} \equiv r_b / r_t$), the 1D fit to $\lambda_{du}$ profiling over the remaining couplings 
yields (when restricted to positive values of the couplings) $\lambda_{du} \simeq 1.05^{+0.35}_{-0.28}$, 
suggesting that the coupling to down-type quarks is not significantly enhanced over the coupling to up-type quarks.

Of course, the lack of evidence for enhanced bottom couplings may be reconciled with non-decoupling $D$-terms 
by raising the masses of the additional supersymmetric Higgs scalars. However this separation might come at the 
price of increasing the tree-level tuning in the Higgs potential unless a theoretically motivated mechanism is introduced to  explain why $|m^2_{H_d}| \gg |m^2_{H_u}|$.\footnote{For particular examples of such models see e.g.~\cite{Csaki:2008sr,Craig:2011yk}.} From a 
pragmatic perspective, this also lessens the prospects for discovering additional Higgs scalars at the LHC if the 
couplings of the recently-discovered resonance prove largely SM-like.

\section{Non-decoupling $D$-terms with chiral Higgses }
\label{sec:chiral}
The undesirable relation between vector-like $D$-terms and enhanced coupling to down-type quarks suggests a natural alternative to raise the Higgs mass: $D$-terms arising from Higgses charged under different gauge groups at high energies,  which we will refer to as ``chiral Higgses".
As before, consider a gauge group at high energies of the form
$SU(3)_c\times SU(2)_1\times SU(2)_2 \times U(1)_1\times U(1)_2$, broken to the Standard Model
around 10 TeV. However, instead of charging the Higgses under the same group, as in the
previous section, consider charging the Higgses under \emph{different groups}, namely
\beq \label{chiralcharges}
H_u \sim (2_{1/2}, 1_0), \ \ \ \ H_d \sim (1_0, 2_{-1/2})~.
\eeq
This seemingly innocent variation on the theme of non-decoupling $D$-terms has far-reaching consequences.
Several immediate implications of this scenario are:
\begin{enumerate}
\item This model is necessarily a model of flavor, as gauge invariance necessarily forbids a subset of MSSM Yukawa
couplings. These forbidden Yukawa couplings may be induced via irrelevant operators, which naturally provides a
rudimentary flavor structure. The particular flavor structure depends on the distribution of MSSM matter fields
between the two groups, which in turn is subject to constraints from anomaly cancelation.

\item A conventional MSSM $\mu$-term is forbidden by gauge invariance.
This is a desirable feature, since it automatically avoids the usual coincidence problem of the $\mu$ scale in the
MSSM. Using the link fields defined in~\eqref{eq:links}, the leading contribution to $\mu$ comes from
the gauge invariant coupling
\beq \label{eq:muterm}
W = \lambda \tilde \chi H_u H_d
\eeq
which leads to an effective $\mu$-term when $\tilde \chi $ gets a VEV:
$\mu_{eff} = \lambda \langle \tilde \chi \rangle$. This provides a reasonable $\mu$ term if
 $\lambda \lesssim 0.1$.

\item It is no longer transparent that custodial symmetry is preserved, and one must check that corrections to the
$\rho$-parameter are under control.
\end{enumerate}

Two first two implications were addressed in~\cite{Craig:2011yk}, and we briefly
revisit them
here. The minimal way to assign the charges of the MSSM matter fields such that all the gauge anomalies cancel out
is to take advantage of the fact that the gauge quantum numbers of the lepton doublet $L$ and $H_d$ in the MSSM are
identical. Therefore we can simply cancel the gauge anomalies by charging a single $L$ under
$SU(2)_1 \times U(1)_1 $,  when all other fields of the same generation are charged under $SU(2)_2 \times U(1)_2$.
In~\cite{Craig:2011yk} we considered the second generation as a ``split'' one, while the entire third generation was
charged under $SU(2)_1 \times U(1)_1$ and the entire first generation was charged under $SU(2)_2 \times U(1)_2$.
This gave us a reasonable flavor structure in the limit $\tan \beta\sim \cO(10^{2}\ldots 10^{4})$, which we later
justify in subsection~\ref{subsec:largetb}. However, this charge assignment breaks full $SU(5)$ multiplets and
 therefore abandons hope of sensible gauge coupling unification. There is, however, a simple alternative that can preserve (accelerated) unification~\cite{ArkaniHamed:2001vr} and gives rise to an effective theory with ``chiral" Higgses, obtained by ``integrating in'' the fields required to the generate Yukawa couplings that are otherwise forbidden by gauge invariance.

\subsection{A simple model}
\label{subseq:simpmodel}

The easiest way to obtain the chiral set of low-energy degrees of freedom with a sensible theory of flavor is to start in the UV with a pair of
Higgs doublets at each $SU(2) \times U(1)$ -- let's call them $A_u, A_d$ charged under
$G_1 \equiv SU(2)_1 \times U(1)_1$, and $B_u, B_d$ charged under $G_2 \equiv SU(2)_2 \times U(1)_2$;
each pair has the quantum numbers of Higgs fields $H_u, H_d$ and some GUT-scale mechanics has eliminated the triplet
counterparts. We also need to assign SM fermions to either $G_1$ or $G_2$. A particularly good choice is to charge the
third generation under $G_1$ and the first two generations under $G_2$; this gives an order-unity top yukawa coupling and
suppressed Yukawa couplings for the lighter generations, though typically the bottom yukawa coupling is suppressed.
Another equally-valid possibility is to charge all three generations under $G_1$; this does not furnish a theory of flavor,
but in this respect it is no worse off than the MSSM. We defer a detailed discussion of the various possibilities for later; the immediate consequences for the Higgs sector are insensitive to this choice.

The possible marginal or relevant superpotential terms allowed by the symmetries involving only the Higgs fields 
are:\footnote{One might object that the presence of $\mu_A, \mu_B$ reintroduces the $\mu$-problem. However, this is far from a $\mu$ problem in the conventional sense; we require only that $\mu_A, \mu_B$ are smaller than $\lambda' \langle \chi \rangle$, and may be arbitrarily small modulo the consequences for flavor.}
\begin{eqnarray}
W \supset \mu_A A_u A_d + \mu_B B_u B_d + \lambda \tilde \chi A_u B_d + \lambda'  \chi A_d B_u ~.
\end{eqnarray}
Now in the limit $\lambda' \langle \chi \rangle \gg \lambda \langle \tilde \chi \rangle, \mu_A, \mu_B$ we may simply integrate out $A_d, B_u$ at tree level, leaving only a low-energy spectrum with $A_u \equiv H_u$,  $B_d \equiv H_d$, and the low-energy superpotential
\beq
W = \left(\lambda \langle \tilde \chi \rangle - \frac{\mu_A \mu_B}{\lambda' \langle \chi \rangle} \right) H_u H_d + \dots
\eeq
Down-type fermion bilinears charged under $G_1$ with tree-level couplings to $A_d$ inherit couplings to
$H_d$ suppressed by $\mu_B/\lambda' \langle \chi \rangle$, while up-type fermion bilinears charged under
$G_2$ with tree-level couplings to $B_u$ inherit couplings to $H_u$ suppressed
by $\mu_A / \lambda' \langle \chi \rangle$. Thus all Yukawa couplings and mixing angles may be generated
without additional matter. This theory is manifestly anomaly-free above the scale
$\lambda' \langle  \chi \rangle$ and below the scale $\lambda \langle \tilde \chi \rangle$. In the
intermediate regime, anomalies are compensated by Wess-Zumino terms in the usual way~\cite{Skiba:2002nx}.
At high energies, if matter fields are distributed in unified multiplets then both groups possess the necessary
matter content to preserve the low-energy supersymmetric prediction of (accelerated) gauge coupling
unification~\cite{Craig:2012hc}. Thus this simple model for chiral $D$-terms possesses all the features
necessary for an anomaly-free theory of flavor, compatible with gauge-coupling unification. Many other variations are possible.

Now we may consider the non-decoupling $D$ terms in the low-energy effective theory. The salient features are most apparent if we work in terms of the effective K\"{a}hler potential, which (for general abelian or non-abelian groups $G_1 \times G_2$) at low energies takes the form
\begin{eqnarray}
K = H_u^\dag e^{g_1 V_1} H_u + H_d^\dag e^{g_2 V_2} H_d + \dots
\end{eqnarray}
in the basis where fundamental and anti-fundamental representations have the same generators. Now after higgsing, the massless vector field $V_D$ and the massive vector field $V_H$ are given by
\begin{eqnarray}
V_D = \frac{g_1 V_2 + g_2 V_1}{\sqrt{g_1^2 + g_2^2}} \hspace{2cm} V_H =
\frac{-g_1 V_1 + g_2 V_2}{\sqrt{g_1^2 + g_2^2}}
\end{eqnarray}

Then we can expand the Kahler potential to leading order in $V_H$ to find
\begin{eqnarray}
K \to H_u^\dag e^{g_D V_D} H_u + H_d^\dag e^{g_D V_D} H_d - \frac{g_D g_A}{g_B} H_u^\dag V_H H_u + \frac{g_D g_B}{g_A} H_d^\dag V_H H_d + \dots
\end{eqnarray}
Integrating out $V_H$ with a spurion for the soft mass term leaves us with an effective action of the form
\begin{eqnarray} \label{eq:asymD}
K_{eff} & = & H_u^\dag e^{g_D V_D} H_u + H_d^\dag e^{g_D V_D} H_d + \\ \nonumber
&& g_D^2 \frac{M_s^2 \theta^4}{M_V^2 + M_s^2} \left(\frac{g_A}{g_B} H_u^\dag T_D^a H_u - \frac{g_B}{g_A} H_d^\dag T_D^a H_d\right)^2 + \ldots
\end{eqnarray}
The relative sign between $H_u$ and $H_d$ in the correction term
arises due to the relative sign in the coupling of the heavy vector superfield $V_H$ to fields charged under $G_1$ and $G_2$.

\subsection{Spectrum and Higgs couplings in the moderate $\tan \beta$ regime}

The contribution of~(\ref{eq:asymD}) for the gauge groups $SU(2)_1 \times SU(2)_2 \times U(1)_1 \times U(1)_2$, with the charge assignments (\ref{chiralcharges}), leads to a potential of the form
\beqa
\Delta V & = &\frac{\xi}{2}\left( \omega |H_u^0|^2 + \omega |H_u^+|^2 +
\frac{1}{\omega } |H_d^0|^2 + \frac{1}{\omega} |H_d^-|^2 \right)^2 + \\
\nonumber && \frac{\xi'}{2}\left( \omega' |H_u^0|^2 + \omega' |H_u^+|^2 +
\frac{1}{\omega' } |H_d^0|^2 + \frac{1}{\omega'} |H_d^-|^2 \right)^2 -\\
\nonumber && 2\xi  \left|H_u^+ (H_d^0)^* + H_u^0 (H_d^-)^* \right|^2
\eeqa
where for simplicity we have defined
\beq\label{eq:xiomega}
\xi \equiv \frac{g^2}{4} \frac{M_s^2}{M_V^2 + M_s^2}, \ \ \ \  \omega \equiv \frac{g_1}{g_2}
\eeq
for the $SU(2)$ contributions, and analogously $\xi \to \xi', \omega \to \omega'$ for the
hypercharge contribution. In the vector case we would have had, e.g., $\Delta = \frac{4}{g^2} \omega^2 \xi$, but in this case it is useful to separate the two factors. Note that in the limit of infinitely large supersymmetry breaking,
the maximal values are $\xi \to \frac{g^2}{4}, \xi' \to \frac{g'^2}{4}$.

This new contribution modifies the usual tree-level relation between the $\beta $ angle and the
soft terms, which now takes the form
\beq\label{eq:sin2bexact}
\sin 2 \beta = \frac{2 B\mu }{ 2|\mu^2| + |m_u|^2 + |m_d|^2 + \xi v^2 f(\omega, \beta)}
\eeq
where
\beq
f(\omega, \beta ) \equiv \left( \omega+ \frac{1}{\omega} \right)
\left( \omega \sin^2 \beta + \frac{\cos^2 \beta }{\omega}\right).
\eeq

The tree-level tadpole conditions also give a modified relation between $m_Z$ and the the soft parameters, namely
\beqa
m_Z^2 & = &\frac{|m_{H_d}^2 - m_{H_u}^2|}{\sqrt{1 - \sin^2(2 \beta)}} - m_{H_u}^2 - m_{H_d}^2 -
2 |\mu|^2 + \\ \nonumber
&&\frac{2 \xi v^2}{\cos(2 \beta)} \left(\omega \sin^2 \beta +
\frac{\cos^2 \beta}{\omega} \right)\left(\omega \sin^2 \beta - \frac{\cos^2 \beta}{\omega} \right)
\eeqa
noting that $\cos(2 \beta) < 0$ for $m_{H_u}^2 < m_{H_d}^2$, as is the case here. At large $\tan \beta$ this becomes
\beq
m_Z^2 + 2 (\xi  \omega^2 + \xi' \omega'^2) v^2  \approx -2 (|\mu|^2 +m_{H_u}^2) + \dots
\eeq
which already indicates that the Higgs mass will accumulate a new positive contribution from the $D$-term corrections.

The pseudoscalar mass in this case is given by $m_A^2 = 2B\mu /\sin 2\beta$,
as in the MSSM, but now the relation~\eqref{eq:sin2bexact} implies a modification of the usual MSSM
expression for $m_A$ in terms of the Higgs soft masses:
\beqa
m_A^2 = |\mu|^2 + m_u^2 + m_d^2 + \xi v^2 \left( \omega+ \frac{1}{\omega} \right) \left( \omega \sin^2 \beta +
\frac{\cos^2 \beta }{\omega} \right) \\ \nonumber +  \xi' v^2 \left( \omega'+ \frac{1}{\omega'} \right)
\left( \omega' \sin^2 \beta + \frac{\cos^2 \beta }{\omega'} \right) ~.
\eeqa
The charged Higgs mass is also deformed relative to the usual MSSM value by a negative contribution
independent of $g_1, g_2$:
\beq
m_{H^+}^2 = m_A^2 + m_W^2 - 2 \xi v^2 ~.
\eeq
Note that this correction is independent of $g_1, g_2$ and brings the charged Higgses {\it closer} to the
pseudoscalar in mass relative to the MSSM. If anything, this improves the fidelity of the custodial $SU(2)$ triplet formed
by $H^\pm, A$ and reduces contributions to the $\rho$ parameter in the low-energy effective theory. However, it does not reverse the ordering
of the spectrum; in the limit of maximal supersymmetry breaking, it predicts exact degeneracy $m_{H^\pm} = m_A$. There are, of 
course, contributions to precision electroweak observables coming from heavy states~\cite{Chivukula:2003wj, Craig:2012hc, Cheung:2012zq}, 
but these may be rendered sufficiently small for the mass scales of interest here.

The  tree-level CP-even Higgs masses are given by
{\footnotesize
\beqa
&m_{H,h}^2 & =  \frac{1}{2} \Biggl( m_A^2 + m_Z^2 + 2\xi v^2 \left( \omega^2 s_\beta^2 + \frac{c_\beta^2 }{\omega^2}\right) + 2\xi' v^2 \left( \omega'^2 s_\beta^2 + \frac{c_\beta^2 }{\omega'^2}\right) \pm \\ \nonumber
&& \sqrt{\left( m_A^2 c_{2 \beta} - m_Z^2 c_{2\beta} + 2\xi v^2(\omega^2 s^2_\beta + \frac{c_\beta^2}{\omega^2}) + 2\xi' v^2(\omega'^2 s^2_\beta + \frac{c_\beta^2}{\omega'^2}) \right)^2 + s^2_{2\beta}
(m_A^2 + m_Z^2 - 2\xi v^2)^2}\  \Biggr),
 \eeqa
}
which we may more usefully expand in the limit $\cos \beta \ll 1 $ and $\xi \ll 1$. To the leading order in $1/ \tan \beta$ and $\xi$, the tree-level masses are simply
\beqa
m_h^2 & = & m_Z^2 + 2 \left( \xi \omega^2 + \xi' {\omega'}^2 \right) v^2 + \cO \left( \frac{1}{\tan^2 \beta} \right) \\
m_H^2 & = & m_A^2 + \cO \left( \frac{1}{\tan^2 \beta} \right).
\eeqa
Thus chiral $D$-terms persist in raising the tree-level prediction for the lightest CP-even Higgs mass, and may
comfortably accommodate a Higgs around 125 GeV without reliance upon large radiative corrections.

\begin{figure}[t] 
   \centering
   \includegraphics[width=2.9in]{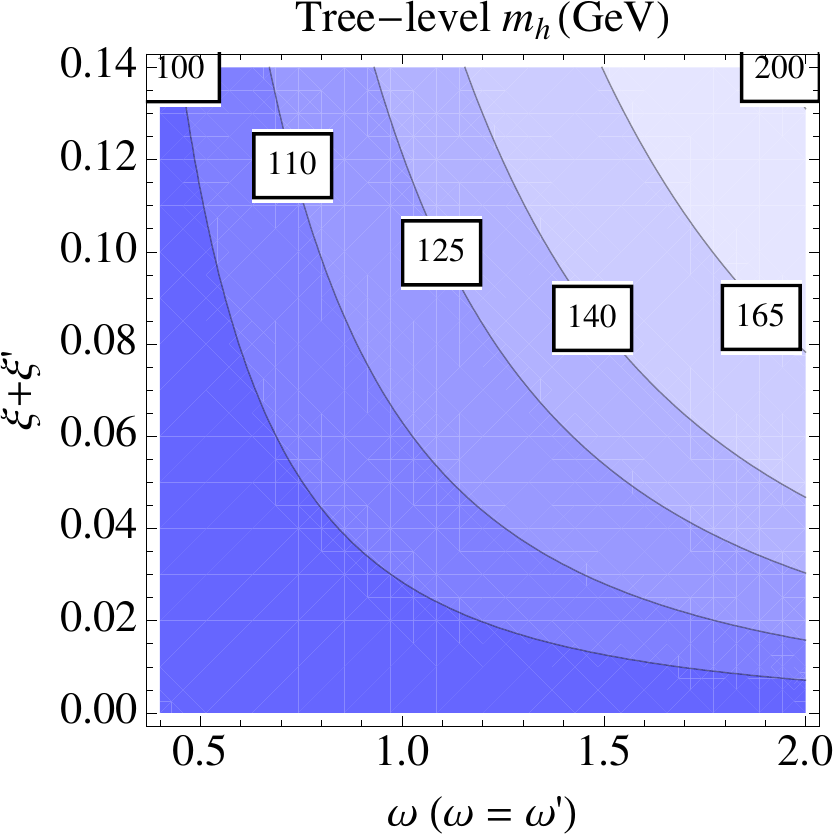} 
    \includegraphics[width=2.93in]{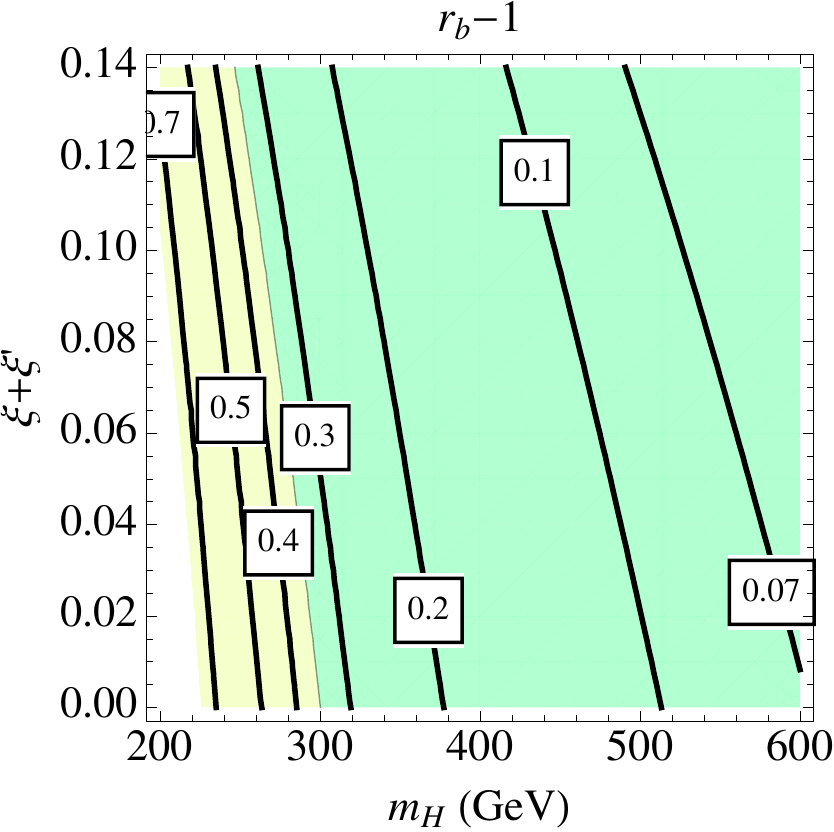} 
   \caption{Left: The tree-level prediction for the Higgs mass as a function of the parameters $\omega$ (taking $\omega = \omega'$) and $\xi + \xi'$. Additional radiative contributions from stop loops further raise the Higgs mass, but are not required to accommodate a Higgs mass around 125 GeV. Right: The enhancement of the bottom coupling $r_b - 1$ above the Standard Model value as a function of the heavy CP-even Higgs mass $m_H$ and the coupling $\xi+\xi'$. The green region contains the $+ 1 \sigma$ fit to $r_b$ using the most recent CMS results; yellow contains the $+2 \sigma$ fit. Points in the white region lie more than $2 \sigma$ outside the CMS fit to $r_b$. Note that the tree-level Higgs mass and couplings are parametrically liberated in this case.}
   \label{fig:chiral}
\end{figure}

In this respect, chiral $D$-terms do not differ significantly from conventional vector-like $D$-terms.
However, they have strikingly different implications for the Higgs couplings. In particular, the correction
from chiral $D$-terms to down-type quark and lepton couplings is now, in the limit of $m_h \ll m_H$ and
$50\gtrsim \tan \beta \gg 1$,
\beq
r_{b/\tau} = \left(1 + \frac{m_h^2}{m_H^2} \right) \left( 1 + \frac{(g^2 + g'^2 - 4 \xi - 4 \xi') v^2}{2 (m_H^2 - m_h^2)} \right).
\eeq
Note the {\it negative} correction to the MSSM contribution. This result is independent of $g_1, g_2$ and, in the limit of
maximal SUSY breaking, reduces precisely to $r_{b/\tau} \simeq 1 + m_h^2 / m_H^2$, entirely 
undoing the effects of
the MSSM quartic couplings. While the corrections from chiral $D$-terms cannot directly suppress the coupling to
bottom quarks below the SM value, they essentially undo the MSSM preference for bottom enhancement. This raises the prospect that, if the couplings of the observed Higgs prove Standard Model-like, there may still be room for additional supersymmetric Higgs scalars at low energies.

\subsection{Higgs couplings in the uplifted regime}
\label{subsec:largetb}
Thus far we have focused on the situation at moderate $\tan \beta$, where tree-level Yukawa couplings dominate. However, as we have pointed out, it is likely that the $B\mu$ term vanishes at the scale $\langle \chi \rangle$ at which the extended gauge symmetry is higgsed. In this case the dominant contribution to the $B\mu$ term comes from the wino loop:
\beq 
B\mu \approx -\frac{3\alpha_2^2}{2\pi} \mu m_{\tilde W} \ln \left( 
\frac{\langle \chi \rangle}{m_{\tilde W}} \right) + \ldots
\eeq 
This leads to the hierarchy $B\mu \ll \mu^2 \sim |m_{H_u}^2|$ and large values of $\tan \beta $. As we pointed 
out in~\cite{Craig:2011yk}, depending on the exact value of the wino mass, $\tan \beta $ in this case can vary
between 40 and $\mathcal{O}\left(10^4 \right)$. While $\tan \beta \sim \cO(10)$ is definitely a viable possibility -- in which case our analysis
of couplings in the previous section is valid -- the situation changes in the ``uplifted'' regime of  $\tan \beta \gtrsim \cO(100)$.    

At such large values of $\tan \beta$, the vev of $H_d$ is too small to provide masses for the down type quarks and the dominant contributions instead come from  ``wrong Yukawas" 
\beq \label{eq:wrongYuk}
\cL \supset \hat Y_b H_u^\dagger Q b^c + \hat Y_{\tau} H_u^\dagger L \tau^c,
\eeq 
which are produced at the loop level. This scenario
was analyzed in~\cite{Borzumati:1999sp, Dobrescu:2010mk} and it was found that in order to get sufficient quark masses one needs
the ``right" bottom and tau Yukawa couplings to be $\cO(1)$, numerically exceeding the top Yukawa. This means 
that these Yukawa couplings reach Landau poles well below the GUT scale. On top of that, this particular part 
of parameter space is subject to severe flavor physics constraints, especially from $B \to \tau \nu$ and 
$B_s \to \mu^+ \mu^- $. These constraints (as well as the issue of Laundau poles for the Yukawa couplings) were
carefully studied in~\cite{Altmannshofer:2010zt}.  Recently improved measurements of $B \to \tau \nu$ and 
$B_s \to \mu^+ \mu^- $~\cite{LHCB:Bstomu} exacerbate constraints on SUSY at large $\tan \beta$. While detailed 
investigation of the numerical consequences of these new constraints is beyond the scope of this work, they merit mentioning before turning to a discussion of Higgs couplings.

Including such ``wrong'' Yukawa couplings, the relative enhancement of bottom couplings in a 2HDM (in an expansion in $1/\tan \beta$) takes the form \cite{Blum:2012ii}:
\beq
r_b = \frac{m_H^2 }{m_H^2 - m_h^2} \left( 1- \frac{1}{1 + \hat Y_b \tan \beta }\left( 
\frac{2 (\lambda_3 + \lambda_5) v^2 }{m_H^2 - m_h^2} - \frac{2 \lambda_7 v^2 \tan \beta}{m_H^2} 
+ \frac{m_h^2}{m_H^2} \hat Y_b \tan \beta\right) \right)~, 
\eeq  
where the couplings $\lambda_3,\ \lambda_7$ and $\hat Y_b$ are defined in Eqs.~\eqref{eq:2hdm}
and~\eqref{eq:wrongYuk}. Until now we have neglected the last two terms in this expression simply because they arise 
at one loop in SUSY models, and so are typically suppressed compared to the $\lambda_3$ term. However,
when $\tan \beta \sim \cO(100)$ we have $\lambda_3 \sim \lambda_7 \tan \beta \sim \hat Y_b \tan \beta$, and 
all the terms in this expression are equally  important.\footnote{Note also that as expected all the new 
contributions are suppressed by $m_H^2$, signaling that in general deviations are small, since one usually
gets very large $\tan \beta$ for $|m_{H_d}^2| \gg |m_{H_u}^2|$. 
} 

Interestingly, the MSSM (as well as the chiral Higgs extension of the MSSM) does not have any firm prediction for the 
signs of $\lambda_7$ and $\hat Y_b$. Both these terms arise at the loop level. MSSM contributions 
to $\lambda_7$ have been calculated in~\cite{Carena:1995bx} and (as expected) are determined by interference 
between three different diagrams; the sign of the result depends on the relative phase between the $\mu$-term 
and the $A$-terms (in our language, it depends on the phase of $\lambda$ in Eq.~\eqref{eq:muterm}). The situation is similar for the correction from $\hat Y_b$, which depends on the relative phases of soft parameters running in the loop. Therefore we conclude that the sign of new contributions to $\Gamma(h \to b \bar b)$ and 
$\Gamma(h \to \tau^+ \tau^-)$  is generally unfixed in the regime of very large $\tan \beta $, allowing for either positive and negative deviation depending on the size and importance of the $\lambda_7$ and $\hat Y_b$ contributions. 

Finally we comment briefly on the $h \to \gamma \gamma $ rate. Preliminary measurements of Higgs BRs by ATLAS and CMS
suggested some deviations from the value predicted by the SM, although these deviations are not
statistically significant and have shrunk with increasing data. However, it is still  interesting to ask whether this particular model has any natural explanation for an enhanced diphoton rate. 
Since we are in the large $\tan \beta $ regime, it is natural to have a light and highly mixed stau. Note that the 
dominant contribution to stau mixing is proportional $Y_\tau \mu$, and $Y_\tau$ is relatively large already by 
$\tan \beta \sim 50 $.  As it was pointed out in~\cite{Carena:2011aa} this can be a source of
 $h \to \gamma \gamma$ enhancement through new loop contributions to the $h \gamma \gamma$ coupling. Moreover, the enhancement need not be as large as in the MSSM, since contributions to the tree-level potential from chiral $D$-terms already undo much of the MSSM preference for bottom enhancement (and hence diphoton BR suppression).  

\subsection{Flavor and sflavor}

Thus far we have remained entirely agnostic as to the distribution of MSSM matter superfields between gauge groups in the UV. This distribution may explain the broad features of Standard Model flavor and/or lead to a natural supersymmetric spectrum, depending on how SUSY breaking is mediated to both gauge groups.  For completeness, we comment briefly on the possibilities:

\begin{itemize}
\item The ``flavorless'' scenario: All MSSM matter superfields are charged under $G_1$. The Standard Model flavor hierarchy is explained through some external physics; the theory is no worse off than the MSSM from the perspective of flavor. The value of $\tan \beta$ depends on how SUSY breaking is communicated:
\begin{enumerate}
\item SUSY breaking is mediated uniformly to both groups via gauge mediation with messengers charged under $G_1$ and $G_2$. Sfermion masses have the usual gauge-mediated spectrum and  $m_{H_d}^2$ is of the order of $m_{H_u}^2$, so that $\tan \beta$ is moderate.
\item SUSY breaking is mediated preferentially to $G_2$ via gauge mediation with messengers charged under $G_2$. Scalars charged under $G_1$ obtain mass via gaugino mediation, with a hierarchy relative to scalars charged under $G_2$ whose size depends on how much the gaugino masses are suppressed. Sfermion masses have a gaugino-mediated spectrum. However, $m_{H_d}^2$ may either be of the same order as $m_{H_u}^2$, in which case $\tan \beta$ is moderate, or it may be much larger, in which case $\tan \beta$ is large and the theory approaches the uplifted limit.
\end{enumerate}
\item The ``flavorful'' scenario: Third-generation MSSM matter superfields are charged under $G_1$, while first- and second-generation matter superfields are charged under $G_2$. This potentially can explain the broad features of the Standard Model flavor hierarchy without large hierarchies in dimensionless couplings, but only under certain circumstances depending on the communication of SUSY breaking:
\begin{enumerate}
\item SUSY breaking is mediated uniformly to both groups via gauge mediation with messengers charged under $G_1$ and $G_2$. Sfermion masses have approximately the usual gauge-mediated spectrum, but third-generation scalar masses may differ somewhat from first- and second-generation masses if the messenger numbers or couplings differ, leading to a mild ``natural SUSY'' hierarchy. Once again, $m_{H_d}^2$ is of the order of $m_{H_u}^2$, so $\tan \beta$ is moderate. However, in this case the tree-level flavor texture generated by marginal and irrelevant operators is poorly aligned with the observed pattern of fermion masses and mixings, so a suitable theory of flavor requires large hierarchies in dimensionless Yukawa couplings.
\item SUSY breaking is  mediated preferentially to $G_2$ via gauge mediation with messengers charged under $G_2$. Scalars charged under $G_1$ obtain mass via gaugino mediation, with a hierarchy relative to scalars charged under $G_2$ whose size depends on how much the gaugino masses are suppressed. If gaugino masses are unsuppressed, there is a mild hierarchy between first/second- and third-generation scalar masses, $\tan \beta$ is moderate, and  a suitable theory of flavor requires large hierarchies in dimensionless Yukawa couplings. If gaugino masses are significantly suppressed, there is a large hierarchy between first/second and third-generation scalar mases, $\tan \beta$ is large, and the uplifted limit is attained. Now the loop-induced Yukawa couplings provide a relatively good fit to the observed pattern of fermion masses and mixings, and no hierarchies in dimensionless Yukawa couplings are required.
\end{enumerate}
\end{itemize}

Thus there are various viable configurations of MSSM matter superfields and messengers of gauge mediation, some of which may explain the broad features of the SM flavor hierarchy. There is typically a correlation between the separation of first/second- and third-generation soft masses and the value of $\tan \beta$; theories with moderate $\tan \beta$ typically have at most a mild, $\mathcal{O}(few)$ hierarchy in the soft masses, while those in the uplifted regime with large $\tan \beta$ have a significant hierarchy between soft masses of the first two generations and the third generation.


\section{Conclusions and Outlook}
\label{sec:conclusions}

A Higgs at 125 GeV poses a sharp challenge for minimal supersymmetric extensions of the Standard Model and suggests either significant fine-tuning, large threshold corrections, or new contributions to the Higgs potential. Such new contributions, if present, may leave their imprint upon Higgs couplings. In the case of $D$-term contributions to the Higgs potential from extended gauge symmetries, the alteration of Higgs couplings typically enhances the partial width to bottom quarks relative to the Standard Model value. This is at variance with current fits to the Higgs couplings and suggests, if the Higgs ultimately proves Standard Model-like, that additional states in the Higgs sector must be significantly heavier than the observed resonance. In this work we have constructed a simple and explicit model for $D$-term corrections to the Higgs potential that instead favors Standard Model-like values for couplings to down-type quarks. This sensibly accommodates the observed Higgs mass and improves the prospects for achieving Standard Model-like decay rates (or perhaps an enhanced diphoton rate) even if the additional states in the Higgs sector are not decoupled in mass. These ``chiral'' $D$-terms automatically imply an approximate theory of flavor and a natural supersymmetric spectrum, as well as additional states that may be within distant reach at the LHC or a future hadron collider.


\acknowledgments{We are grateful to Alex Azatov, Kfir Blum, Spencer Chang, Jamison Galloway, Michele Redi, 
David Shih, and Scott Thomas for useful discussions. NC is supported in part by  the DOE under grant 
DE-FG02-96ER40959, the NSF under grant PHY-0907744, and the Institute for Advanced Study. 
AK is supported by NSF grant No. PHY-0855591. 
We acknowledge the hospitality of the Aspen Center for Physics, supported by the 
National Science Foundation Grant No. PHY-1066293, where parts of this work were completed.
AK also  thanks the New High Energy Theory Center at Rutgers University for hospitality.} \\

{\bf Note added:} While this paper was being prepared for submission, \cite{Huo:2012tw} and \cite{D'Agnolo:2012mj} appeared. Both references discuss gauge extensions of the MSSM with vector-like charge assignments for the Higgs doublets; \cite{Huo:2012tw}  emphasizes the implications for Higgs couplings from loop-level contributions, while \cite{D'Agnolo:2012mj} reiterates the tension between vector-like models and current fits to Higgs couplings.

\bibliography{lit}
\bibliographystyle{jhep}

\end{document}